\documentclass[useAMS,usenatbib]{mn2e}

\DeclareTextSymbol{\degre}{OT1}{23}
\usepackage{graphicx}

\title[Mass ratios of 20 new SB2s]{Masses of the components of SB2 binaries observed with Gaia.\\ I. Selection of the sample and mass ratios of 20 new SB2s discovered with Sophie\footnotemark[1]\thanks{based on observations performed
at the Observatoire de Haute--Provence (CNRS), France}}
\author[J.-L. Halbwachs et al.]{J.-L. Halbwachs$^{1}$\thanks{E-mail:
jean-louis.halbwachs@astro.unistra.fr}, F. Arenou$^{2}$, D. Pourbaix$^{3}$, B. Famaey$^{1}$, P. Guillout$^{1}$,
  \newauthor
Y. Lebreton$^{2,4}$, J.-B. Salomon$^{1}$, L. Tal-Or$^{5}$, R. Ibata$^{1}$ and T. Mazeh$^{5}$\\
$^{1}$Observatoire astronomique de Strasbourg, CNRS UMR 7550,
   11 rue de l'Universit\'{e}, F--67000 Strasbourg, France\\
$^{2}$Observatoire de Paris, GEPI, UMR 8111, F--92195 Meudon, France\\
$^{3}$FNRS, Institut d'Astronomie et d'Astrophysique, Universit\'{e} Libre de Bruxelles, boulevard du Triomphe, 1050 Bruxelles, Belgium\\
$^{4}$Institut de Physique de Rennes, Universit\'e de Rennes 1, CNRS UMR 6251, F-35042 Rennes, France\\
$^{5}$School of Physics and Astronomy, Tel Aviv University, Tel Aviv 69978, Israel}
\begin{document}

\date{Accepted 2014 September 3.  Received 2014 September 3; in original form 2014 June 30}

\pagerange{\pageref{firstpage}--\pageref{lastpage}} \pubyear{2014}

\maketitle

\label{firstpage}

\begin{abstract}
In anticipation of the Gaia astrometric mission, a large sample of spectroscopic binaries is being observed
since 2010 with the Sophie spectrograph at the Haute--Provence Observatory. Our aim is to derive the orbital elements of double-lined
spectroscopic binaries (SB2s) with an accuracy sufficient to finally obtain the masses of the components 
with relative errors as small as 1~\% when the astrometric measurements of Gaia are taken into account. 
Simultaneously, the luminosities of the components in the Gaia photometric band $G$ will also be obtained.
Our observation program started with 200 SBs, including 152 systems that were only known as single-lined.
Thanks to the high efficiency of the Sophie spectrograph, an additional component was found for 25 SBs.
After rejection of 5 multiple systems, 20 new SB2s were retained, including 8 binaries with evolved primary, and their mass ratios 
were derived. Our final sample contains 68 SB2s, including 2 late-type giants and 10 other evolved stars.
\end{abstract}



\section{Introduction}

Mass is the most crucial input in stellar internal structure modelling.
It predominantly influences the luminosity of a star and, therefore, its lifetime.
Similarly, masses, ages, and initial helium abundances of stars are mandatory prerequisites to study processes with different time scales, including the formation and evolution of planetary systems, the dynamical and chemical evolution of our Galaxy and even the distant unresolved galaxies. 
Precise and accurate outputs from models of stellar internal structure and evolution are essential in many astrophysical studies. In particular, the age and helium content of stars are not accessible to observers and have to be inferred from stellar modelling. 
Unfortunately, the mass of a star can generally be independently determined only when the star belongs to a binary system.
The knowledge of the mass of stars in a binary system, together with the assumption that the components have the same age and initial chemical composition, allows one to determine the age and the initial helium content of the system and therefore to characterize the structure and evolutionary stage of the components
\citep[see e.g.][]{LebFerLe,Ribas,FerVaVic}.
Such modelling provides insights on the physical processes governing the structure of the stars and gives constraints on the free physical parameters of the models, provided the masses are known with great accuracy \citep{Lebreton05}. Therefore, modelling stars with extremely accurate masses (at the 1\% level), for different ranges of masses, would allow one to firmly anchor the models of the more loosely constrained single stars.

At present, accurate masses are still rare:
\citet{TorresAnG10} have listed the non-interacting systems with masses more accurate than 3\%. They found 95 eclipsing binaries 
(EBs) and 23 astrometric binaries (ABs).
Therefore, the EBs are now the most common providers of masses, and, in addition, the analysis of their light curves can also provide accurate radius estimates. However, they are rare among binary stars, since their orbital plane must be oriented close to edge-on. As a consequence, there may be no or few EBs that are suitable for the study of some classes of stars. In contrast, observations of ABs are only limited by instrumental performance, and the rate of these systems will grow when the next generation of astrometric instruments comes on-line. The Gaia astrometric satellite should be the cause of this change.

Thanks to Gaia,
astrometric orbits will be obtained for several systems which are already known as spectroscopic binaries (SB). When the radial velocities of both components of an SB will be measured, i.e. for double-lined SBs (SB2s), the products ${\cal M}_1 \sin^3 i$ and ${\cal M}_2 \sin^3 i$ may be derived from the orbital elements; therefore, when the inclination $i$ of the orbit will be derived from Gaia observations, the masses of the components, ${\cal M}_1$ and ${\cal M}_2$, will be derived too.
In addition, since the semi-major axis of an astrometric orbit is related to the luminosity ratio of the components, the individual magnitudes in the Gaia $G$ band will also be obtained.
A similar method was applied by \citet{2000IAUS..200P.135A} or \citet{RenFu10}, using the astrometric measurements of the Hipparcos satellite; nevertheless, in the latter case, the results were significant for 13 systems only, and the minimum uncertainty of the masses was around 3\%. As early as 1999, \citet{Baltic99} have shown that the situation will be quite different with Gaia, due to the very high accuracy of this satellite. However, to fully take benefit of the exquisite accuracy of Gaia, the elements of the spectroscopic orbits must also be very precise. It was then decided to select
known SBs, and to observe them with a spectrograph dedicated to the search of extrasolar planets, namely the Sophie spectrograph. Masses with relative errors smaller than 1\% can then be obtained.

The selection of the sample of SBs is presented in Section~\ref{sec:sample} hereafter. This sample initially included a majority of single-lined binaries (SB1), since
the detection of the spectral lines of the secondary component was expected for a fraction of them. The observation program that was carried out at the Haute--Provence Observatory since 2010 is presented in Section~\ref{sec:observations}, and the results concerning new components are detailed in Section~\ref{sec:additional}.


\section{Selection of the sample}
\label{sec:sample}

\subsection{Choice of the criteria}
\label{sec:problem}
The main source of spectroscopic binaries with known orbits is the SB9 Catalogue \citep{SB9}, which is regularly updated and accessible on-line\footnote{http://sb9.astro.ulb.ac.be/}. The current version of the catalogue
contains the orbital elements of 3208 SBs, and also, generally, their old radial velocity (RV) measurements. We included also in the selection process a few orbits that were still
not in the SB9 at that time (in 2009): the orbits selected in \citet{JLH-Gliese} and the orbits published in \citet{CPM}.

Despite the quality of Gaia measurements, it will not be possible to derive accurate masses for all these stars, and it was necessary to select those which could really be attributed masses
with a 1\% accuracy. Schematically, this accuracy may be achieved if
the RVs are sufficiently accurate to provide ${\cal M}_1 \sin^3 i$ and ${\cal M}_2 \sin^3 i$ with errors smaller than 1\%, and if Gaia will also provide $\sin^3 i$ with errors below this limit. The minimum masses are calculated from the period, $P$, the eccentricity, $e$, and the semi-amplitudes of the RV of the components, $K_1$ and $K_2$, using the following relation:

\begin{equation}
{\cal M}_{1,2} \sin^3 i = (K_1 + K_2)^2 K_{2,1} (1 - e^2)^{3/2} P \times 1.036 \; 10^{-7}
\label{eq:msini}
\end{equation}

\noindent
where $P$ is in days, $K_{2,1}$ in km/s and the masses in solar units. If the orbital elements are derived from old measurements completed with recent ones, we may expect that the
period will not significantly contribute to the error budget of the minimum mass. The main contribution comes from the semi-amplitudes. 
With velocimeters providing RV with errors usually between 0.3 and 0.5 km/s, the old orbits provide minimum masses with uncertainties equal to a few percent. If the errors are divided by 10 or more, the orbits that will be obtained will have minimum masses with the requested accuracy. Obtaining accurate masses depends then only on the derivation of $\sin^3 i$ from the Gaia astrometric measurements. To estimate if the relative uncertainty of $\sin^3 i$ will be below 1\%, it is necessary to simulate the astrometric orbit observed with Gaia. 
The first step is the derivation of the input parameters of the astrometric orbit, as explained hereafter.


\subsection{Parameters of the astrometric orbits}
\label{sec:parSB}
The spectroscopic orbital elements of an SB lead to several parameters of the astrometric orbit. They include the period, the eccentricity, the epoch and the longitude of periastron.
The missing parameters are the position angle of the ascending node, $\Omega$, the semi-major axis of the photocentric orbit, $a_0$, and the inclination of the orbit. The position angle $\Omega$ was
randomly generated, but it could have been arbitrarily fixed as well, since this would not have changed the results of the simulations. The choice of the method of derivation of the inclination is tricky: For an SB2, it is tempting to derive $i$ from the value of ${\cal M}_1 \sin^3 i$ coming from the spectroscopic elements, assuming a value of ${\cal M}_1$ corresponding to
the spectral type of the star. However, this would lead to an overestimation of $\sin i$ when the actual mass is larger than the mass given by the calibration. 
Our simulations have shown
that the uncertainty of the mass of the components is much smaller when the inclination is large, i.e. when the system is oriented edge-on, than when it is small (near pole-on orientation).
Therefore, deriving $i$ from ${\cal M}_1 \sin^3 i$ would lead to selecting preferably SB2s with masses larger than the value coming from the spectral type. In order to avoid this bias, we
preferred to generate $i$ from the $\sin i$ distribution, although this distribution refers to all binaries as a whole. This method was also applied to the SB1s.

Once the inclination is known, the semi-major axis of the photocentric orbit may be derived from the following equation:

\begin{equation}
a_0 = \frac{a_1 \sin i}{\sin i} \left( 1 - \beta\frac{1+q}{q} \right) \varpi
\label{eq:a0}
\end{equation}

\noindent
where $a_1 \sin i$, in astronomical units, is derived from the elements of the spectroscopic orbit, and $\varpi$ is the parallax, in the same unit as $a_0$.  The other terms are the mass ratio, $q={\cal M}_2/{\cal M}_1$, and the luminosity fraction in the $G$ photometric band of Gaia, $\beta={\cal L}_2/({\cal L}_1+{\cal L}_2)$. 
The mass ratio is known for the SB2s, and also for some SB1s for which it was derived from Hipparcos observations \citep{JLH-Gliese}. For the other SB1s,
$q$ is derived
from the mass function, assuming the inclination obtained as above and deriving the primary mass from the spectral type.
When the magnitudes of both components were known, $\beta$ was derived from the difference of
magnitudes.
Otherwise, it was derived from the mass ratio.
In both cases, it was necessary to derive $\Delta G$, the difference of the magnitudes of the components in the
$G$ band. From the relations given by \citet{Landolt} for the main-sequence stars and from a preliminary
calibration of $G$, it was derived that $\Delta G = -7.3 \log q$, and, therefore, $\Delta G = 0.73 \Delta V$.
It is worth noticing that a new relation between the color indices $(G-V)$ and $(B-V)$ was published
later by \citet{Jordi}, and that the slope of the relation ``absolute G magnitude vs $\log {\cal M}$'' 
is now around -9.5 rather than -7.3, as we assumed. Therefore, it is possible that a few SBs were
discarded in the selection process because the semi-major axis of the orbit of the photocentre
was slightly underestimated.

Once the parameters of the astrometric orbit were generated, they were used to simulate Gaia observations.


\subsection{Simulation of Gaia observations}
\label{sec:simGaia}

We used the model of Gaia observations described in \citet{perspective}. This model is crude as it does not reproduce the relation
of the scanning law with the coordinates of the star.
It leads however, on average, to the same uncertainties as the accurate model. We remind the reader that
an astrometric observation of Gaia consists in the measurement of the abscissa of the
photocentre of the binary star along a scanning axis. The position of the photocentre was derived from the parameters of the
astrometric orbit, and also from the coordinates, the parallax, and the proper motion of the barycentre of the binary.
An instrumental error was generated randomly, and it was added to the abscissa in order to produce an astrometric measurement.
A complete astrometric solution was derived from the simulated measurements. This solution included the parallax, the proper motion,
the semi-major axis, the orbital inclination,
but not the period, the eccentricity, the epoch, and the longitude of the periastron, since they are supposed to be fixed to
their actual values thanks to the known spectroscopic orbit. The error of $\sin^3 i$ was thus derived for each simulated orbit. Ten thousand
orbits, based on 10\,000 values of the inclination, were generated for each SB, and the proportion of solutions with relative errors of $\sin^3 i$
smaller than 1\% was finally obtained. When this proportion was larger than 20\%, the SB was selected.

The procedure described above was applied to the SBs of our input list. As mentioned above, SB1s were also included in the selection, but with a restriction:
in practice, the secondary spectrum is visible with Sophie only when the mass ratio is larger than around 0.5. Therefore, the SB1s were considered for selection only when the
minimum mass ratio, as estimated from the mass function, was larger than 0.3. Otherwise, the mass ratio could be as large as 0.5, but only if the inclination of
the orbit is small. Since the orbits with small inclinations have large errors for $\sin^3 i$, it was better to discard them.

Only the stars fainter than $6^{\rm m}$ were selected, since, at the time the sample was chosen, Gaia was not expected to make usable observations of brighter stars. 
The stars with declinations below -5\degre were also rejected, since they could not be easily observed from the South of France.
After rejection of the systems known to be triple, a selection of 200 SBs with a probability better than 20\% to get components masses more accurate than 1\% was finally obtained. It contained 152 SB1s and 48 SB2s.


\section{The observations}
\label{sec:observations}
The observation program began in 2010, and it is still on-going. We are using the T193 telescope of the Haute--Provence Observatory, 
with the Sophie spectrograph in high resolution mode. The minimum signal-to-noise ratio is 50. When this would lead
to observing time shorter than 3 minutes, it is increased to 100 or to 150. Since the orbital elements of the stars are already known, ephemerides
are computed before each run, in order to observe with higher priority the stars where the components have RV sufficiently different to be
accurately measured. 

From 2010 to 2013, our program totalled 30 nights of observations, distributed over 16 observing runs during which 727
spectra have been acquired on the 200 target stars.
The Sophie pipeline immediately derives the cross-correlation function (CCF) of each spectrum with a mask function chosen by the observer
to match the stellar spectral type.
The range of derivation of the CCF is chosen in order to include the correlation dips of both components, when they are visible. Thanks to this
facility, the orbital elements of the stars were verified, and, whenever necessary, they were updated for the next run.

The CCFs of the SB1 were inspected in order to find the correlation dip of the secondary component. When the companion
was not visible, despite a favorable orbital phase, the star was flagged as confirmed SB1, and set aside. Otherwise, it was flagged as a new SB2. 


\section{The additional components discovered with Sophie}
\label{sec:additional}


\subsection{Double-lined binaries and multiple systems}

The dip of an additional component is visible on the CCFs of 25 stars which were previously classified as SB1. Among these stars, it appears that 5 are clearly
at least triple, and that 20 are SB2. The multiple systems are  BD +13 331, HIP 35664, HD 115588, HD 149240 and HD 207739. 
More information about these stars is provided in the notes, Section~\ref{sec:notes}. 


\subsection{Derivation of the radial velocities of the components of the new SB2s}
\label{sec:RV}

\begin{figure}
\includegraphics[clip=,height=2.2 in]{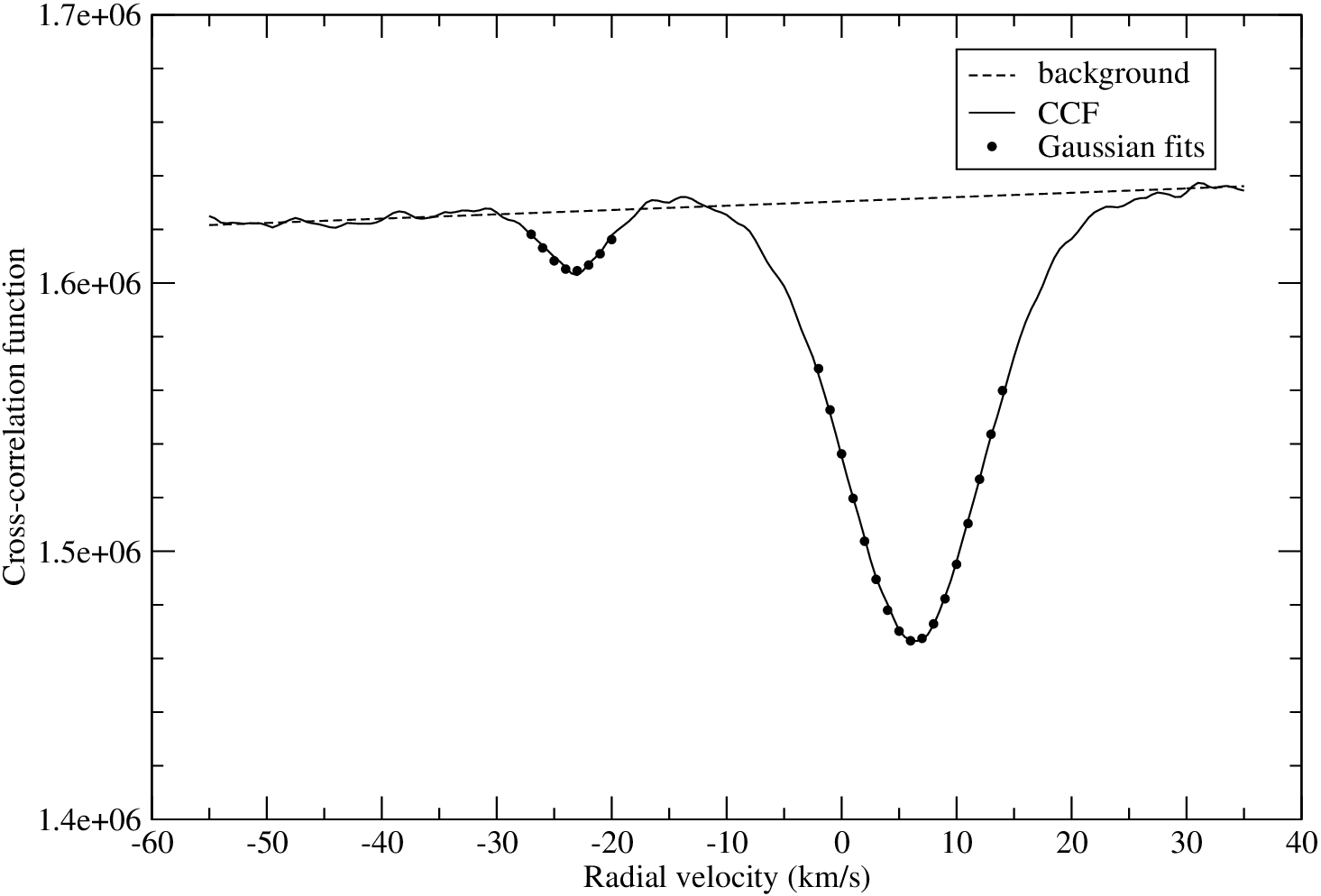}
 \caption{The CCF of a spectrum of HIP 12472, obtained with the G2 mask. The secondary component is clearly visible on the left of the primary.
The dashed line is the background.
The bold dots are normal distributions fitted to the CCF
on 1.4$\sigma$ intervals around the minima. The velocities of the components are -23.2 and 6.32~km/s.
}
\label{fig:CCF}
\end{figure}

The RVs of both components were derived from the CCFs, as described hereafter. This method is not as sophisticated as the two-dimensional correlation algorithm 
TODCOR
\citep{zucker94, zucker04}, but it is sufficient for a preliminary estimation based on a small number of observations.
It was already mentioned in Section~\ref{sec:observations} that the Sophie pipeline provides the CCF of each spectrum with a mask.
An example is presented in Fig.~\ref{fig:CCF}. The first step of the reduction consists in fitting a linear function to the background. The normalized slope, i.e. the slope divided by the estimation of the background for $RV=0$, is then fixed, and two normal distributions are fitted to the CCF around the minima corresponding to
the components. The RVs of both components are thus derived. The model applied to the CCF contains 8 parameters: the normalized slope of the background, the
value of the background for $RV=0$, and, for each dip, the normalized depth, the standard deviation and the position of the minimum. When several
spectra of the same SB are available, the calculation is improved by considering all the CCFs at the same time. Many parameters are then common to all of the CCFs: the normalized slope of the background is set to its mean value, and,
in addition to the background level for $RV=0$, only the positions of the dips, i.e. the RV of the components, are varying from one CCF to another. 
In order to compensate the errors coming from the difference between the model and the actual CCFs,
the RV errors obtained from this computation were corrected in order to have $F_2=0$, where $F_2$ is the estimator of the goodness-of-fit defined in \citet{Kendall},
as:

\begin{equation}\label{def:F2}
F_2= \left( \frac{9\nu}{2} \right)^{1/2} \left[ \left( \frac{\chi^2}{\nu} \right)^{1/3}+{\frac{2}{9 \nu}}-1 \right]
\end{equation} 

where $\nu$ is the number of degrees of freedom and $\chi^2$ is the weighted
sum of the squares of the differences between the predicted and the observed
values, normalized with respect to their uncertainties. When the predicted values
are obtained through a linear model, $F_2$ follows the normal distribution
{\cal N}$(0,1)$. When non--linear models are used, but when the errors are small
in comparison to the measurements, as hereafter, the model is approximately linear around the
solution, and $F_2$ follows also 
{\cal N}$(0,1)$.

The RVs derived with this method depend on two parameters, which are the parts of the CCF used to fit the normal profiles of the dips, and the normalized slope of the background.
The effects of these parameters are discussed hereafter. First of all, the normal distribution is a very good model to describe the CCF near the minimum of a dip, but, when one moves away from the minimum, the real curve suddenly diverges from the model: in reality, the transition from the bell shape to the flat background is not so smooth as in the normal
distribution. Therefore, the normal distribution must be fitted only on a restricted range around the minimum of the dip. The extension of this range depends on the star: for
a late-type dwarf with a narrow dip of around 3 km/s, it may be more than twice the standard deviation of the dip, $\sigma$. In contrast, for some early-type stars with correlation dips 
as wide as 40 km/s, it does not exceed $0.5 \sigma$. It is worth
noticing that the error generated by the discrepancy between the normal model and the actual shape of the dip is very small when single stars are considered. For double stars, errors may come from the contamination from the dip of the companion. This effect is avoided by restricting again the range around the minima, and by discarding the CCFs with dips too close one to another. For this reason, for the computations hereafter, the adopted range never exceeds $1.4 \sigma$.
The other effect is the slope of the background. The actual background is not perfectly a line, as assumed in the model, and, even when it is fitted to several CCFs, the
normalized slope assumed in the calculation is a bit arbitrary. However, changing the slope essentially results in shifting the velocities of both components by constant values,
which depend on the standard deviations of the dips. Therefore, the orbital elements and the mass ratio remain the same when they are derived introducing among the parameters a constant offset between the velocities of the components.


\subsection{The mass ratios of the new SB2s from preliminary orbital solutions}
\label{sec:q_orbital}

The mass ratio of an SB2 is usually obtained from the orbital elements, through the equation $q_K={\cal M}_2/{\cal M}_1=K_1/K_2$.
Using this formula requires the derivation of all the orbital elements of the SB2, a computation which is not possible when 
only the few spectra obtained with Sophie are taken into account. However, when the old measurements supplied by the SB9 catalogue
are added, this calculation is possible with only 3 spectra giving the RVs of both components, when they are sufficiently
different: the measurements of the RV of the primary components lead to a verification, and, possibly, a correction of the periastron epoch and of the period; the measurements of the secondary RV lead to the offset between the Sophie measurements of the components, and to an estimation of $K_2$. 

The mass ratios in the column ``$q_K=K_1/K_2$'' of Table~\ref{tab:q} were thus obtained. As in the previous step, the uncertainties coming from the least-square computation were corrected in order to have $F_2=0$.

No result is given for HIP 101452, since 
the measurements used to derive the old orbit were never published.


\subsection{Derivation of the mass ratio directly from the CCF}
\label{sec:q_RV}

The mass ratio of an SB2 may be derived from the Sophie spectra alone, without computing the orbital elements,  
simply assuming instead that the
RV of the secondary component is varying symmetrically to that of the primary. 
It obeys the relation $v_2=v_\gamma - (v_1-v_\gamma)/q$,
where $v_\gamma$ is the RV of the barycentre and $v_1$ is the RV of the primary component. Therefore, $v_2$ may be written as:

\begin{equation}
v_2 = v_{02}  - \frac{v_1}{q}
\label{eq:v2}
\end{equation}

The term $v_{02}$ is constant, since $v_{02} = v_\gamma ( 1 + 1/q)$. It comes from equation~(\ref{eq:v2}) that $v_2$ is a linear
function of $v_1$ with the slope $-1/q$. This conclusion is valid even when the RV of both components are each shifted with a constant
systematic error. This results in adding to $v_{02}$ the difference of the errors, but this does not affect the estimation of $q$.

\begin{figure}
\includegraphics[clip=,height=2.2 in]{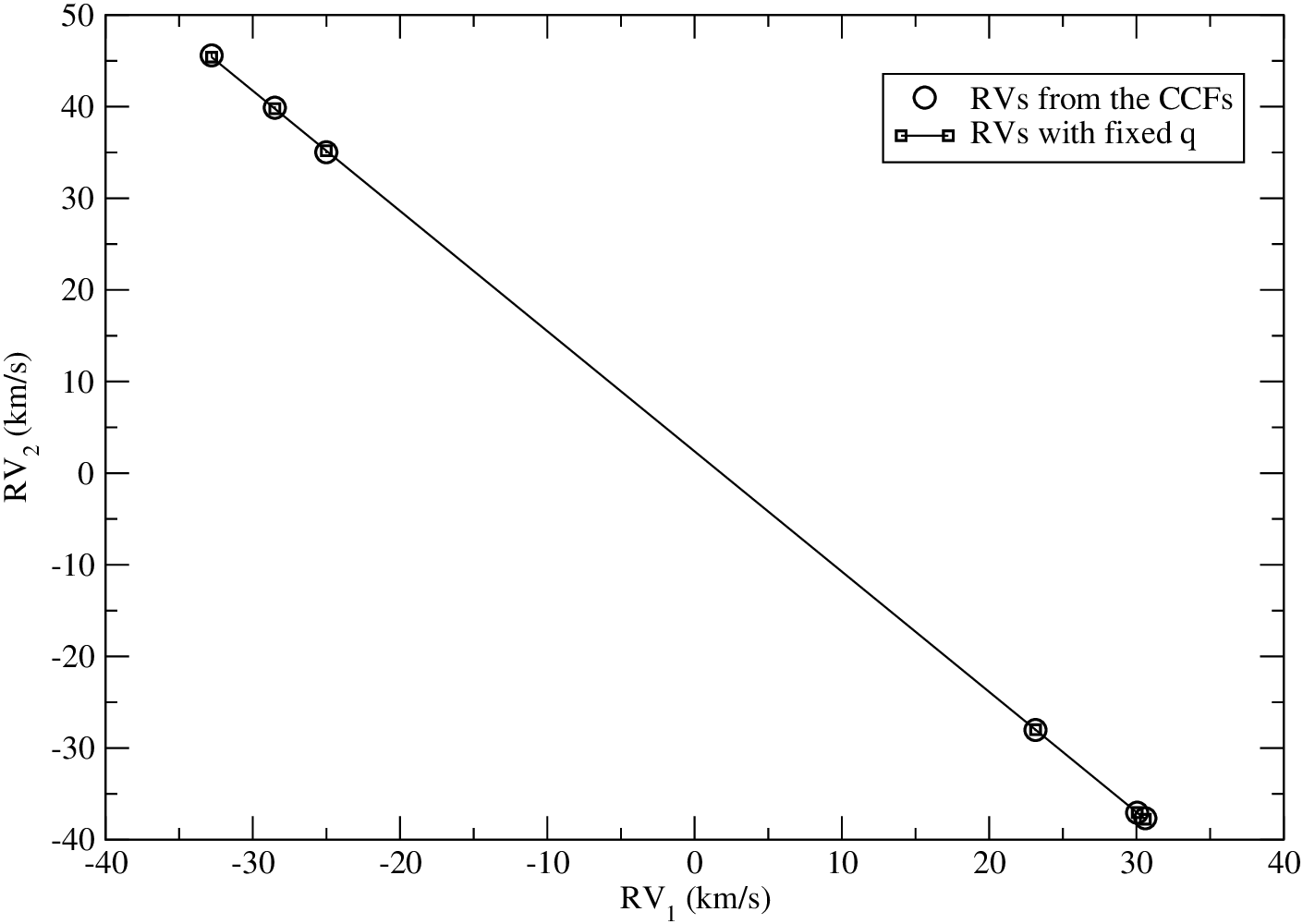}
 \caption{Derivation of the mass ratio of HIP 7143. The RV measurements of the secondary component are plotted as a function of the RV of the primary.
The circles are the RV measurements directly obtained from the CCFs, as explained in Section~\ref{sec:q_orbital}. The
squares are the measurements obtained from the CCFs, deducing the secondary RVs from the mass ratio as explained in Section~\ref{sec:q_RV}. The mass ratio is $q=-1/s$, where $s$ is the slope of the line connecting the points.
}
\label{fig:V1V2}
\end{figure}

In principle, in order to obtain the mass ratio, it is then sufficient to consider the velocities of the components at two different
phases of the orbit, corresponding to two different values of $(v_1, v_2)$. This method is illustrated in 
Fig.~\ref{fig:V1V2}.

Again, the mass ratio is derived from a unique computation based on all the CCFs selected to derive $K_2$ above. The model used to derive the RV is modified, thanks to
equation~\ref{eq:v2}. The parameters $v_{02}$ and $q$ are then common to all the CCFs, and only $v_1$ is changing. Again,
the RVs of each component are affected by
a bias depending on the spectral type, but this bias does not affect the
evaluation of the mass ratio, but only that of $v_{02}$.

The mass ratios
obtained are listed in Table~\ref{tab:q}, in the column ``$q_\Delta=\Delta V_1/\Delta V_2$''. As for the other estimations of $q$, the
uncertainties are corrected in order to have $F2=0$.
Although they were obtained from the same CCFs, the mass ratios are not exactly equal to those obtained in Section~\ref{sec:q_orbital}.
The difference comes from the methods: $q_K$ is derived assuming a Keplerian motion, and taking into account the old measurements.
In contrast, $q_\Delta$ is derived from the CCF alone, assuming a linear relation between the RVs of the components. 
This linear relation is not a consequence of Kepler's laws, but only of the principle of Galilean relativity: the
barycentre of the system is moving with a constant velocity.

A comparison with
column $q_K$ shows that both methods give nearly the same results; this confirms their reliability.

\begin{table*}
 \caption{The mass ratios of the new SB2s.}
\label{tab:q}
\begin{tabular}{@{}lrrccrlllc}
\hline
Identification & Sp.T. & SNR & mask & $l/\sigma$ & $N$ & $q_K=K_1/K_2$ & References & $q_\Delta=\Delta V_1/\Delta V_2$ & rem \\ 
\hline
HIP 626 & G5 & 150 & K0 & 1.4 & 5 & $0.606 \pm 0.072$ & \cite{Massarotti}  & $0.606 \pm 0.034$ & *  \\ 
HIP 7134 & G5 & 100 & K0 & 1.4 & 5 & $0.7059 \pm 0.0037$ & \cite{GrifEm}  & $0.7061 \pm 0.0032$ & * \\
HIP 7143 & G5 & 100 & K0 & 1.4 & 6 & $0.7624 \pm 0.0015$ & \cite{Heard}  & $0.7622 \pm 0.0011$ & * \\ 
HIP 12472 & F5 & 100 & G2 & 1.4 & 4 & $0.6369 \pm 0.0022$ & \cite{CarGriGi}  & $0.6390 \pm 0.0026$ & \\ 
HIP 13791 & F8V & 50 & G2 & 1.4 & 6 & $0.591 \pm 0.018$ & \cite{Imbert}  &  $0.5844 \pm 0.0063$ & \\ 
HIP 24035 & G5 & 50 & K0 & 1.4 & 3 & $0.665 \pm 0.037$ & \cite{CPM} & $0.666 \pm 0.037$ & \\ 
HIP 25160 & A2m & 100 & G2 & 1.0 & 5 & $0.869 \pm 0.038$ & \cite{Imbert} &  $0.910 \pm 0.018$ & * \\ 
HIP 29982 & G5III & 150 & K0 & 1.4 & 3 & $0.7016 \pm 0.0093$ & \cite{Griffin86}  & $0.7020 \pm 0.0092$ & * \\ 
HIP 48895 & F3V & 100 & G2 & 0.5 & 4 & $0.875 \pm 0.018$ & \cite{Griffin06}  & $0.864 \pm 0.015$& * \\
HIP 61727 & F3V & 100 & G2 & 1.4 & 6 & $0.63801 \pm 0.00088$ & \cite{CPM}  & $0.63819 \pm 0.00077$& *\\
HIP 61732 & G5 & 50 & K0 & 1.4 & 5 & $0.7071 \pm 0.0030$ & \cite{CPM}  & $0.7068 \pm 0.0018$& *\\
HD 110106 & K3V & 50 & K5 & 1.4 & 4 & $0.843 \pm 0.056$ & \cite{CPM}  & $0.849 \pm 0.056$ & * \\ 
HIP 62935 & G5 & 50 & K0 & 1.4 & 7 & $0.6926 \pm 0.0041$ &  \cite{Griffin04} & $0.6914 \pm 0.0040$ &   \\ 
HIP 67195 & F5 & 100 & G2 & 1.0 & 6 & $0.5740 \pm 0.0013$ & \cite{Shajn}  & $0.5748 \pm 0.0010$ & \\ 
HIP 69481 & F8 & 100 & G2 & 1.0 & 6 & $0.513 \pm 0.014$ & \cite{Bakos} & $0.5077 \pm 0.0085$ & *  \\ 
HIP 72706 & K0 & 100 & K5 & 1.4 & 5 & $0.7851 \pm 0.0033$ & \cite{Massarotti}  &  $0.7849 \pm 0.0023$ & *   \\ 
HIP 94371 & G5V & 50 & K0 & 1.4 & 4 & $0.7309 \pm 0.0049$ & \cite{Griffin79}  & $0.7296 \pm 0.0042$ & \\ 
HIP 101452 & A0p & 100 & F0 & 1.0 & 3 & - &  & $0.719 \pm 0.011$ & * \\ 
HIP 110900 & K0V & 100 & K5 & 1.4 & 3 & $0.72 \pm 0.12$ & \cite{Massarotti}   & $0.731 \pm 0.062$ & * \\ 
HIP 114661 & F6 & 50 & G2 & 1.4 & 3 & $0.5013 \pm 0.0032$ & \cite{Latham}  & $0.4995 \pm 0.0036$ &    \\
\hline
\end{tabular}

 \medskip
The column $l/\sigma$ gives the range of the CCF taken into account around each correlation dip, expressed in unit of the standard deviation
of the dip.
$N$ is the number of spectra taken into account to derive the mass ratio, $q$. 
An asterisk in the remark column indicates a note
in Section~\ref{sec:notes}.

\end{table*}

\begin{figure}
\includegraphics[clip=,height=2.2 in]{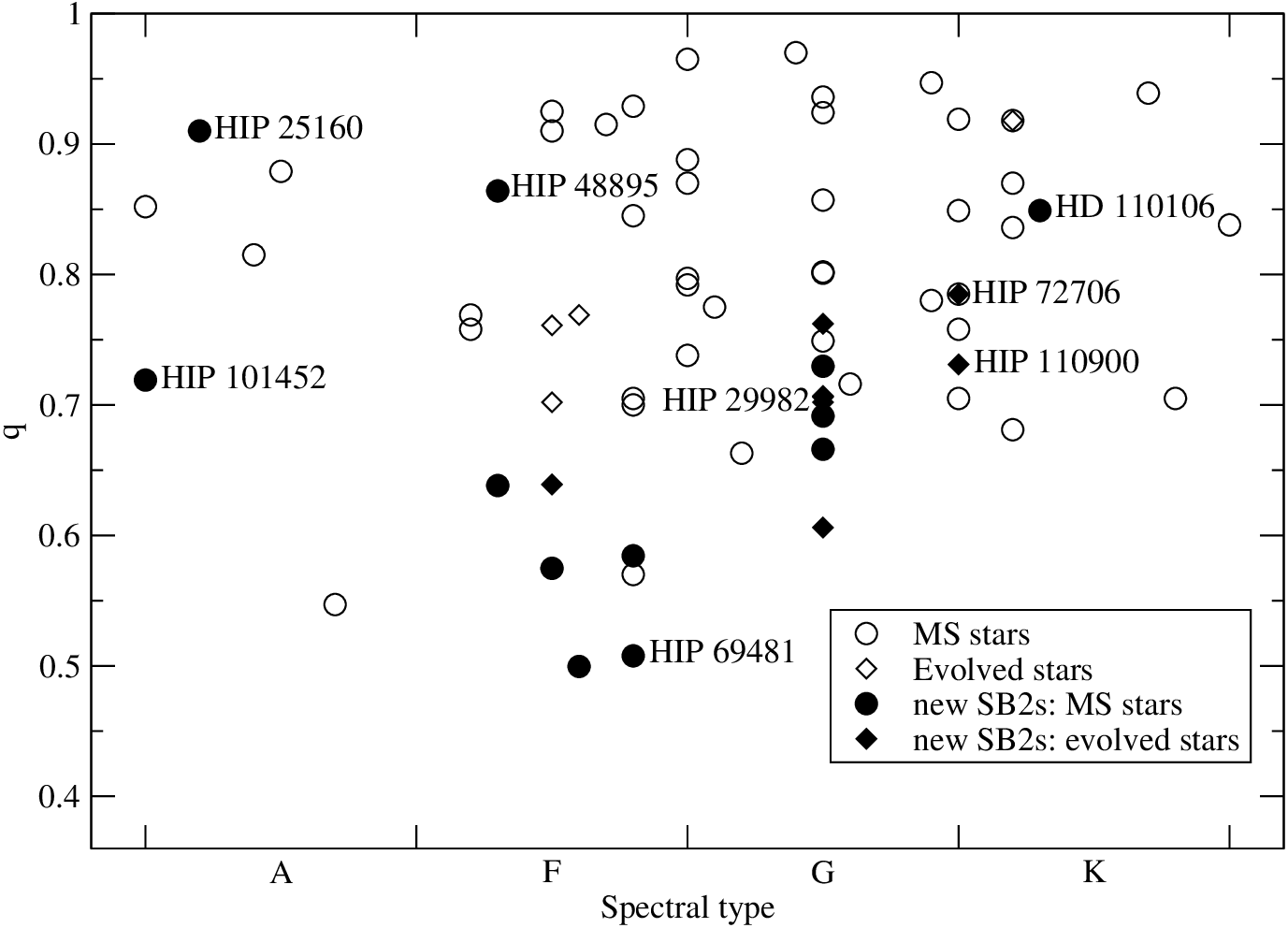}
 \caption{The mass ratios as a function of the primary spectral types, for the SB2s already known (open symbols) as for the new ones (filled symbols). 
The ``evolved stars'' are stars classified as giant or subgiant, or stars more
than 2 magnitudes brighter than the main
sequence. The identifications of some stars discussed in the text are indicated.
}
\label{fig:q-TSp}
\end{figure}

The mass ratios are plotted versus the primary spectral types in Fig.~\ref{fig:q-TSp}. The absolute magnitudes of the Hipparcos stars were
derived from their trigonometric parallax, and the stars more than 2 mag above the main sequence are indicated as ``evolved'' in this figure.
It appears, then, that 8 of the new SB2s are evolved stars, including 7 G-type or K-type stars.

Adding these 20 new SB2 to the 48 which were already known, we obtain a total sample of 68 SB2s with spectral types from A0 to M0. 
This sample contains two late-type giant stars. In addition, when we count 10 stars classified as subgiant, or without known luminosity class but 
more than 2 mag brighter than the main sequence. It is worth noticing that, among the 12 evolved stars thus selected, 8 are new SB2s.
The difference in the proportion of evolved stars among the previous SB2s (4/48) and among the new ones (8/20) cannot
be a random effect: a calculation based on the hypergeometric distribution shows that this hypothesis is rejected at the 0.4 \% level of significance.
It is due to the fact that Sophie makes possible the detection of a faint secondary component.


\subsection{Notes about individual objects}
\label{sec:notes}

\begin{description}
\item[\bf HIP 626.] The mass ratio is rather uncertain, since all the spectra are so blended that the secondary dip is always embedded in a wing
of the primary dip.
The SB1 orbit by \citet{Massarotti} does not fit our measurements. It seems that the period is around 3250 days instead of 1568.
The star is probably evolved, since it is 3.4 mag brighter than a G5 dwarf.

\item[\bf HIP 7134.] A preliminary SB2 orbit was published by \citet{SF2A12}.

\item[\bf HIP 7143.] \citet{Heard} derived an SB1 orbit, but he noticed that some spectral lines were obviously blended. A preliminary SB2 orbit was published by \citet{SF2A12}.

\item[\bf BD +13 331. ] \citet{GrifYoMil} pointed out that the mass function of the SB1 orbit is abnormally large, and suggested that the system is multiple.
A secondary dip is clearly visible in the CCFs of this star, though its motion is not symmetrical to that of
the primary one. The secondary is then itself an SB1.

\item[\bf HIP 25160.] An Am binary with wide correlation dips. The secondary dip is clearly separated from the primary one on two observations only, and the velocities are nearly the same for both epochs. Therefore, it was necessary to take into account CCFs with dips severely blended to derive the mass ratio.

\item[\bf HIP 29982.] An SB2 with a late-type giant primary component. The secondary dip is the smallest among all the new SB2s, and it was discovered thanks to the large SNR (150) applied to this star. The depth is only 3.4~\% of that of the primary dip, and the dip is always contaminated by the side lobes of the primary dip.  

\item[\bf HIP 35664.] Three spectra of this G2 V star were obtained, and, although each of the three CCF exhibit a single dip, its shape is variable, suggesting that the system is triple. The
visible component would be a close pair with a small orbital inclination.

\item[\bf HIP 48895.] An F3 star with wide correlation dips: the standard deviations are around 13 and 40 km/s, respectively, and all our CCFs show the narrow and deep primary dip
emerging from the wide secondary.
\citet{Griffin06} reported that the secondary spectrum was detected by \citet{Albitzky} on photographic plates; he measured the velocities of both components from 4 observations, but he preferred to neglect the secondary measurements in the computation of the orbital elements. We derived the SB2 orbit assuming that, when Griffin measured only one component,
the RV does not refer to the primary component, but to a blend of the correlation dips, and we applied the computation method of \citet{CPM}. 

\item[\bf HIP 61727.] A preliminary SB2 orbit was published by \citet{CPM}.

\item[\bf HIP 61732.] A preliminary SB2 orbit was published by \citet{CPM}.

\item[\bf HD 110106.] The estimation of $q$ is uncertain, since the period is very long (approximately 8 years), and the RV of the components changed little over the year covered by our observations. Nevertheless, the mass ratio is surprisingly large
for a dwarf late-type star. The correlation dips are both narrow, but the depth ratio is small (11~\%), explaining that the secondary dip was 
not detected on the Coravel observations of \citet{CPM}. 
The luminosity class of the primary star is not confirmed with a trigonometric parallax, so it may be a classification error.

\item[\bf HD 115588.] \citet{Griffin07} proposed that the large mass function of the SB1 orbit is due to a secondary component that is double in reality.
We confirm this hypothesis, since, aside from the primary dip, two additional dips are clearly visible on the CCF of a spectrum taken on JD2456034.53872.

\item[\bf HIP 69481.] An F8 star with a very wide correlation dip (25 km/s). The secondary is faint, but it is clearly visible thanks to the narrowness of its
correlation dip (3.5 km/s). $q_K$ was derived using the RV measured by \citet{Bakos}, but discarding the RV obtained from
old low-dispersion spectra that he used also in the derivation of the SB1 orbit.

\item[\bf HIP 72706.] The mass ratio is rather large for a system with a late-type primary component, suggesting that the star is evolved. This hypothesis is supported by the bright absolute magnitude of the system: $3^{\rm m}$.

\item[\bf HD 149240.] \citet{Griffin82} noticed that the correlation dip of the star is not as deep as expected for a K0 star, and that the mass function is excessively large. Our observations confirm that the system is multiple: a deep secondary dip is clearly visible on the CCF of the first spectrum, which was taken on JD2455306.52685. It was not observed again later, even when the primary component had the same RV as at the first observation, but the primary dip was sometimes 
asymmetrical, suggesting that a secondary dip was embedded in a wing.

\item[\bf HIP 101452.] An A0p star with a secondary dip almost as deep as the primary one, but which previously escaped detection due to its
width (about 30 km/s). \citet{Northcott} published an SB1 orbit, but she didn't provide the RV measurements; therefore, it was not possible to
compute $q_K$. 

\item[\bf HD 207739.] An F8 + B binary system studied by \citet{GrifPa}, who pointed out the very large mass function of the SB1 orbit derived from the F-type component. The B component is not visible on the CCFs that we obtained, but the correlation dip is clearly double on one observation. The F-type
component is probably an SB2.

\item[\bf HIP 110900.] Another small secondary dip, close to the detection limit for a star observed with SNR=50. The depth is only 7~\% of that of the primary dip. In addition, the only CCF with the secondary dip on the right side of the primary is severely blended, contributing to the large uncertainty of $q$.
The absolute magnitude of the star is mag 3, indicating that its luminosity class is probably not V, but IV. This would explain the difficulty to detect
the secondary component.

\end{description}


\section{Discussion and conclusion}

We have selected 200 SBs for which it should be possible to derive the masses of the components with an accuracy close to 1~\%, 
merging data of two different kinds: the astrometric measurements of the Gaia satellite, and high-precision RV measurements
of both components.
These stars have been observed for 8 semesters with the T193 telescope of the Haute--Provence Observatory, and, thanks to 
the sensitivity of the Sophie spectrograph, we obtained a final sample of 68 double stars which are now observed in order to
obtain the masses of the components with high accuracy when the astrometric measurements of Gaia will be available.
Thanks to the 20 new SB2s we have found, the number of evolved stars increased from 4 to 12, including 2 late-type giants.
The mass ratios
of the new SB2s are derived from our measurements. The new
SB2s are distributed in 2 categories, according to the correlation dip of the secondary component:

\begin{itemize}

\item
The secondary component is a low--rotation star. 
The majority of the stars are in this category. The secondary component 
was too faint to be visible on
photographic plates of the spectra, or on the CCF obtained with Coravel.
The primary component is an early-F type star, or a star of later type. 
When it is a fast rotator, the narrow correlation dip of the secondary is easily visible when added to the wide primary dip, allowing detection even when the mass ratio is as small as around 0.5 (HIP 69481).
When it is a main-sequence star, the mass ratio is usually between 0.5 and 0.7. 
When it is an evolved star, the mass ratio is larger than 0.7.

\item
Pair of fast rotators. The primary component is an A-type or a F-type star, with a wide correlation dip. The secondary is a star similar to the
primary, with a wide correlation dip. The mass ratio is then rather large ($q > 0.7$), since the secondary component must be bright to be detected. HIP 25160, 48895 and 101452 belong to this category.

\end{itemize}
 
The mass ratios of the new SB2s are listed in Table~\ref{tab:q}, so they are available for any project requiring this information.
For instance, they may be used to derive the semi-major axes of the relative orbits and to select targets for interferometric observations. Another application would be the derivation of the distribution of $q$ in statistical
studies.

We also notice that, despite the small number of measurements and the difficulty to detect the secondary dips, the mass ratios were derived with an accuracy better than 1~\% for 13 of the new SB2. Our observation program will be continued until we will have enough measurements to derive the elements of the
SB orbits from the Sophie RV alone, and this will still require a few years. Nevertheless, the accuracy of the mass ratios already obtained
looks very promising for the final results of the program.

\section*{Acknowledgments}

It is pleasure to thank Dr X. Delfosse for helpful suggestions.
This project was supported by the french INSU-CNRS ``Programme National de Physique Stellaire''
and ``Action Sp\'{e}cifique Gaia''. We are grateful to the staff of the
Haute--Provence Observatory, and especially to Dr F. Bouchy, Dr H. Le Coroller, Dr M. V\'{e}ron, and the night assistants, for their
kind assistance. We used two facilities operated at CDS, Strasbourg, France: the Simbad object database and the VizieR catalogue database.

\bsp

\label{lastpage}

\end{document}